# Why Systems-on-Chip Needs More UML like a Hole in the Head

Stephen J. Mellor, John R. Wolfe, Campbell McCausland
Accelerated Technologies
Embedded Systems Division of Mentor Graphics
Tucson AZ, USA

## Abstract

*Let's be clear from the outset: SoC can most certainly make use of UML; SoC just doesn't need more UML, or even all of it. The advent of model mappings, coupled with marks that indicate which mapping rule to apply, enable a major simplification of the use of UML in SoC.*

## 1. The Need for an SoC Abstract Modeling Language

At the beginning of an SoC project, it is common for the hardware and software teams to work a specification in parallel. Invariably, the two components do not mesh properly. You can't verify understanding until you have something to execute; and once you have something that executes, it costs a lot to change the interface.

Once the prototype runs, it is possible to measure the performance, which may require changing the partition. Partition changes are expensive, and are difficult to do correctly.

Moreover, the system is usually "specified" only in terms of implementation. We need a way to model a system at an appropriate level of abstraction that does not presume an implementation in software or hardware. SystemC and Handel-C are low-level, and presume too much implementation.

## 2. Executable and Translatable UML

The introduction of the Action Semantics enables execution of UML models. The Executable UML profile [1] defines a carefully selected streamlined subset of UML and defines an execution semantics for it.

The essential elements are a set of classes and objects with concurrently executing state machines. State machines communicate only by sending signals. On receipt of a signal, a state machine executes a set of actions that runs to completion before the next signal is processed. The actions in the destination state of the receiver execute *after* the action that sent the signal. This captures desired cause and effect.

A model can be executed independent of implementation. No design details or code need be added, so formal test cases can be executed against the model to verify that requirements have been properly met. Critically, Executable UML is structured to allow developers to model the underlying semantics of a problem without having to worry about whether it is to be implemented in hardware of software.

Them's the rules, but what is really going on is that Executable UML is a concurrent specification language. Rules about synchronization and object data consistency are simply rules for that language, just as in C++ we execute one statement after another and data is accessed one statement at a time. We specify in such a concurrent language so that we may *translate it* onto concurrent, distributed platforms; hardware definition languages; as well as fully synchronous, single tasking environments.

## 3. Marks

Marks describe models but they are not a part of them, rather like sticky notes. A mark is a lightweight, non-intrusive extension to models that captures information required for mappings without polluting those models. Mappings rules are applied to model elements that have been marked to indicate which rule to apply—hardware or software. An example is the mark *isHardware*, which may be associated with some element to be implemented in hardware. This allows for retargeting models to different implementation technologies as they change. Mappings and marks are described in [2].

## 4. Model Mappings

At system construction time, the conceptual objects are mapped to hardware and software. Repeatable mappings are defined that produce compliable text (*e.g*., C, VHDL) according to a *single consistent set of architectural rules*. The mapping rules therefore guarantee that the interfaces are consistent. A model compiler interprets the mapping rules to maintain the desired sequencing specified in the models, which it may do any manner it chooses so long as the defined behavior is preserved.

The result is several text files of two (in this example) types. One is all the C that is to be implemented in software; the other is VHDL. The two halves are known to fit together because the interface was generated. Changing the partition is a matter of changing the placement of the marks.

## 5. A Hole in the Head

Executable UML is a small, but powerful, subset of UML enabling abstract specification of behavior. Mappings enable interface definition in one place, so that consistency is guaranteed. Marks enable late decision making on the partition. That's all we need; we need more UML like a hole in the head.